\documentclass[runningheads]{llncs}
\usepackage[section]{placeins}
\usepackage{url}
\usepackage{amsmath}
\usepackage{amssymb}
\usepackage{multirow}
\usepackage{float}

\usepackage{graphicx}
\usepackage{hyperref}
\begin{document}
\mainmatter              
\title{A Mathematical Approach to Improve Energy-Water Nexus Reliability Using a Novel Multi-Stage Adjustable Fuzzy Robust Approach}
\titlerunning{Energy-water nexus planning under uncertainty}  
%
\author{Afshin Ghassemi\inst{1} \and Michael J Scott\inst{2}
}
\authorrunning{Afshin Ghassemi and Michael J Scott} 
%
\tocauthor{Afshin Ghassemi and Michael J Scott}

\institute{University of Illinois at Chicago, Chicago IL 60607, USA,\\
\email{afshinghassemi@gmail.com}
\and
University of Illinois at Chicago, Chicago IL 60607, USA,\\
\email{mjscott@uic.edu}
}

\maketitle              

\begin{abstract}
A system of a systems approach that analyzes energy and water systems simultaneously is called energy-water nexus. Neglecting the interrelationship between energy and water drives vulnerabilities whereby limits on one resource can cause constraints on the other resource. Power plant energy production directly depends on water availability, and an outage of the power systems will affect the wastewater treatment facility processes. Therefore, it is essential to integrate energy and water planning models. As mathematical energy-water nexus problems are complex, involve many uncertain parameters, and are large-scale, we proposed a novel multi-stage adjustable Fuzzy robust approach that balances the solutions' robustness against the budget-constraints. Scenario-based analysis indicates that the proposed approach generates flexible and robust decisions that avoid excessive costs compared to conservative methods.
\end{abstract}
\section{Introduction}
In mathematical energy-water nexus planning, the aim is to offer a general model that generates energy decisions and water decisions while covering these two complex systems' interconnections.
The energy-water nexus planning has been thoroughly studied via discussions of policies and regulations during the last fifty years \cite{carrillo2009water,mackres2011addressing,park2012californias,tarroja2014evaluating}. The recent research in this area provided more accurate data for energy-water nexus planning (e.g., energy for water and water for energy) \cite{Wolff2004,Wakeel2016,DOE2006,Mielke2010a,Stillwell2011}. The last class of the studies applied mathematical approaches toward energy-water nexus problems, where limited numbers of them propose a general mathematical model or framework for energy-water nexus planning\cite{Yang2016,denooyer2016integrating,SANDERS2015317,santhosh2014real,chen2016urban,WANG2018353}. 
The growing demand for shared resources is predicted to be one of the primary subjects of conflicts in the approaching years, not only on the connected energy-water area but in many other critical areas (see \cite{2015JIEI...11..543E,mohamadi2020nash}).
When studying the energy planning and water planning simultaneously in the general model, the uncertainty is considered in both the supply and demand sides of the energy-water nexus \cite{Hassanzadeh2015,Arandia2015,JU2016184}. Overlooking these randomnesses in systems can undoubtedly lead to infeasible solutions. 
Some studies have analyzed uncertainty for at least part of the energy-water nexus planning \cite{Ghassemi2017,babayan2005least,Weini2013,PARISIO201437,COELHO2016567,SUGANTHI2015585}. Ghassemi \cite{ghassemi2019system} offers a robust framework for their research. One of the main methods to deal with uncertainty is using Fuzzy logic which has vast applications in optimization under uncertainty \cite{pejoo,PEYKANI2019439}. 
This research aims to extend on that method and propose a flexible, multi-stage, Fuzzy robust approach for energy-water nexus planning. The coupling of the Fuzzy method and robust optimization makes it possible to cover diverse types of uncertainties while keeping the system's flexibility. 

The new, flexible, multi-stage, Fuzzy Robust method is described in detail section \ref{section2}. The next section describes the energy-water systems profile and provides the results of the scenario-base analysis section \ref{section3}. The discussion of this study and future research plans are presented in section \ref{section4}.

\section{Optimization Under Uncertainty Using Robust and Fuzzy Methods}\label{section2}
Using the Fuzzy and robust optimization method leads to an increase in the total systems cost. Models that are flexible to generate a range of solutions are more preferred than the traditional single-solution methods. The managers would be able to generate decisions with a certain level of robustness based on their desired criteria, such as the budget constraints.

In energy-water nexus planning, dismissing the possible scenarios’ details will lead to loss of information and irreversible consequences. 

A new two-stage adjustable robust approach is introduced in \cite{ghassemi2019system} that provides a balance between solution robustness and the budget constraints. In this work, the Fuzzy decision models' concepts are integrated into that proposed model, and a new adjustable, multi-stage, robust, Fuzzy approach is proposed. 

\subsection{Fuzzy Logic Integration} \label{test}
The possibility, necessity, and credibility indexes represent the chances of occurrence of Fuzzy events with an optimistic, pessimistic, and moderate view towards planning, respectively. Here, the conversion of Fuzzy chance constraints to their equivalent crisp for a specific confidence level is provided. In this paper, $\tilde{\nu}$ and $\gamma$ are Fuzzy and crisp numbers. It is assumed that the uncertain parameter is a Fuzzy number with trapezoidal distribution that is showed by $\tilde{\nu}\,({{\nu }_{1}},{{\nu }_{2}},{{\nu }_{3}},{{\nu }_{4}})$ while ${{\nu }_{1}}<{{\nu }_{2}}<{{\nu }_{3}}<{{\nu }_{4}}$.

To integrate the Fuzzy logic, a Fuzzy number with trapezoidal distribution of $\tilde{\nu }\,({{\nu }_{1}},{{\nu }_{2}},{{\nu }_{3}},{{\nu }_{4}})$ is presented. 
The triple ($\Psi$, P($\Psi$), $Pos$) represent a possibility space that a universe set $\Psi$ is a non-empty set, covering all possible events and P($\Psi$) be the power set of $\Psi$. 

The credibility measure is defined as the average of the possibility and necessity measures as $Credibility\{.\}=\frac{1}{2}(Possibility\{.\}+Necessity\{.\})$.
\newline
The credibility profile measures are defined as follows:
\begin{itemize}   
\item $Credibility\{\varnothing\}=0$
\item $Necessity\{\Psi\}=1$
\item $if \;\; A\in P(\Psi)\;  \Rightarrow \;  0\le Credibility\{A\}\le 1$
\item $if \;\;{{A}_{i}}\in P(\Psi) \;\; and \;\; Su{{p}_{i}}(Credibility\{{{A}_{i}}\})<0.5  \Rightarrow   Credibility\{ {{\cup }_{i}}{{A}_{i}} \}=Su{{p}_{i}}(Credibility\{{{A}_{i}}\})$
\item $if \;\;A\in P(\Psi)  \Rightarrow   Credibility\{A\}+Credibility\{{{A}^{C}}\}=1                                                   \;\;\;(Self-Duality)$
\item $if\;\; A,B\in P(\Psi) \;\; and \;\; A\subseteq B  \Rightarrow   Credibility\{A\}\le Credibility\{B\}                             \;\;\; \\     (Monotonicity)$
\item$if \;\;A,B\in P(\Psi)  \Rightarrow   Credibility\{A\cup B\}\le Credibility\{A\}+Credibility\{B\}                                    \;\;\;\\ (Subadditivity)$
\item$Possibility\{A\}\ge Credibility\{A\}\ge Necessity\{A\}$
\end{itemize}
Based on the credibility measure, the Eqs. in \ref{9} and \ref{10}, transforms the Fuzzy chance constraints for confidence level of $\alpha$ as:
\begin{equation}\label{9}
Credibility\{\tilde{\nu }\le \gamma \}=\left\{ \begin{aligned}
  & 0, \;\;\;\;\;\;\;\;\;\;\;\;\;\;\;\;\;\;\; if \;\;\;{{\nu }_{1}}\ge \gamma  ; \\ 
 & \frac{\gamma -{{\nu }_{1}}}{2({{\nu }_{2}}-{{\nu }_{1}})},   \;\;\;    if \;\;\; {{\nu }_{1}}\le \gamma \le {{\nu }_{2}} ; \\ 
 & \frac{1}{2}, \;\;\;\;\;\;\;\;\;\;\;\;\;\;\; \;\;\;if \;\;\;{{\nu }_{2}}\le \gamma \le {{\nu }_{3}} ; \\ 
 & \frac{\gamma -2{{\nu }_{3}}+{{\nu }_{4}}}{2({{\nu }_{4}}-{{\nu }_{3}})}, if \;\;\; {{\nu }_{3}}\le \gamma \le {{\nu }_{4}} ; \\ 
 & 1, \;\;\;\;\;\;\;\;\;\;\;\;\;\;\;\;\;\;\; if \;\;\;                 {{\nu }_{4}}\le \gamma  . \\ 
\end{aligned} \right.
\end{equation}
\begin{equation}\label{10}
Credibility\{\tilde{\nu}\ge \gamma \}=\left\{ \begin{aligned}
  & 1,\;\;\;\;\;\;\;\;\;\;\;\;\;\;\;\; \;\; if \;\;\; {{\nu }_{1}}\ge \gamma  ; \\ 
 & \frac{2{{\nu}_{2}}-{{\nu }_{1}}-\gamma }{2({{\nu }_{2}}-{{\nu }_{1}})},       if  \;\;\;{{\nu }_{1}}\le \gamma \le {{\nu }_{2}} ; \\ 
 & \frac{1}{2}, \;\;\;\;\;\;\;\;\;\;\;\;\;\;\;\;\; if  \;\;\;{{\nu }_{2}}\le \gamma \le {{\nu }_{3}} ; \\ 
 & \frac{{{\nu }_{4}}-\gamma }{2({{\nu }_{4}}-{{\nu }_{3}})},    \;\;\; if  \;\;\; {{\nu }_{3}}\le \gamma \le {{\nu }_{4}} ; \\ 
 & 0,\;\;\;\;\;\;\;\;\;\;\;\;\;\;\;\;\; if \;\;\;{{\nu }_{4}}\le \gamma  . \\ 
\end{aligned} \right.
\end{equation}
The Eqs. in \ref{9} and \ref{10} can be summarized as:
\begin{equation}\label{11}
Credibility\{\tilde{\nu }\le \gamma \}\ge \alpha \;\;\; \Leftrightarrow \;\;\; \left\{ \begin{aligned}
  & (2-2\alpha ) {{\nu }_{3}}+(2\alpha -1) {{\nu }_{4}}\le \gamma \;\;\;if \;\;\;\alpha >0.5 ; \\ 
 & (1-2\alpha ) {{\nu }_{1}}+2\alpha  {{\nu }_{2}}\le \gamma \;\;\;\;\;\;\;\;\;\;\;\; if \;\;\;\alpha \le 0.5 . \\ 
\end{aligned} \right.
\end{equation}
\begin{equation}\label{12}
Credibility\{\tilde{\nu }\ge \gamma \}\ge \alpha \;\;\; \Leftrightarrow \;\;\; \left\{ \begin{aligned}
  & (2\alpha -1) {{\nu }_{1}}+(2-2\alpha ) {{\nu }_{2}}\ge \gamma \;\;\; if \;\;\;\alpha >0.5 ; \\ 
 & 2\alpha  {{\nu }_{3}}+(1-2\alpha ) {{\nu }_{4}}\ge \gamma \;\;\;\;\;\;\;\;\;\;\;\; if \;\;\;\alpha \le 0.5 . \\ 
\end{aligned} \right.
\end{equation}
\subsection{The Multi-stage Fuzzy Robust Approach Architecture}\label{section 2.2}
In energy-water nexus planning, the system could face different scenarios, and each of these scenarios has a designated probability of happening that a Fuzzy number is assigned to represent that.
In the multi-stage Fuzzy robust approach, three types of decisions should be processed. The first two are the Fuzzy weights' discrete decisions and then the Fuzzy numbers in the uncertain model. The third decision is related to parameters with continues ranged uncertainty that is calculated by the robust optimization. The third decision is related to parameters with continues ranged uncertainty, which is calculated by the robust optimization.

\begin{figure}[hbt!]
	\begin{center}
			\includegraphics[scale=0.47]{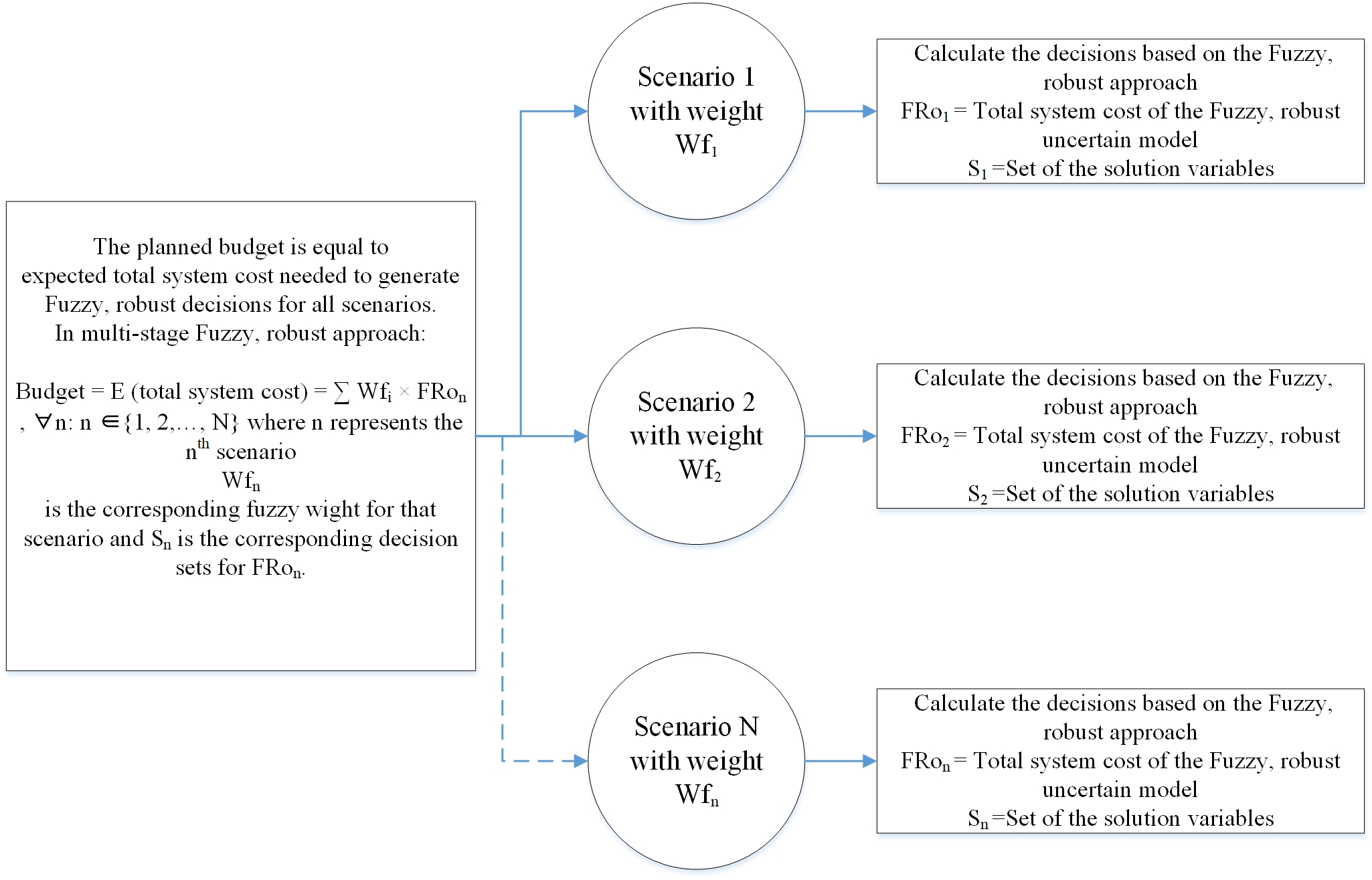} 
		\caption{The architecture of the multi-stage adjustable Fuzzy robust approach to find the expected total system cost}
		\label{figgg}
	\end{center}
\end{figure}

In the robust approach, the first step calculates the total system cost based on the uncertain model. The next step is to convert the weights based on one of the mentioned Fuzzy methods (necessity, possibility, or credibility) to nominal values and then normalize the new weights by dividing the weight for each scenario to the sum of the weights for all possible scenarios. The last step is to find the sum-product of each scenario's weights multiplied by the corresponding total system cost.
Figure~\ref{figgg} illustrates the multi-stage Fuzzy robust approach for energy-water system robust optimization.

\section{Scenario Analysis} \label{section3}
In this section, first, the parameters and the features of the energy-water nexus model are explained. Then, using a large-size example, the proposed model's performance is examined, and the results are investigated.
\subsection{The Energy-water System Profile}
The uncertain energy-water nexus model and the parameters such as power generator information and scenario parameters are the same as the author's previous work \cite{ghassemi2019system} with two main changes. The weights are considered to be represented by trapezoidal Fuzzy numbers. The efficiency parameters are modeled by both robust and Fuzzy methods. The planning horizon is 10$\times$24, and the model generates online hourly decisions. The final model is a Fuzzy robust, multi-period, linear mixed-integer optimization problem. The energy-water nexus model covers water and wastewater extraction, transmission, purification, and all the associated costs. The energy part covers thermal power systems, such as power transmission systems, power generators, and batteries. The objective function includes two categories of penalty costs for unmet demand and untreated wastewater to reduce unwanted results.

\subsection{Numerical Analysis}
In order to demonstrate the benefits of the new multi-stage Fuzzy robust planning method, the expected total system cost is calculated on two levels of \textit{$\mathit{\alpha}={0}$} and \textit{$\mathit{\alpha}={0.5}$}  (\textit{$\mathit{\alpha}$} works as a control parameter for the level of conservativity in Fuzzy methods) using the credibility Fuzzy method. The robust levels are calculated for \textit{$\mathit{\varGamma}={0}$}, \textit{$\mathit{\varGamma}={0.25}$},
\textit{$\mathit{\varGamma}={0.5}$},
\textit{$\mathit{\varGamma}={0.75}$}, 
\textit{$\mathit{\varGamma}={1}$}
(the parameter $\mathit{\varGamma}$ is a control of how conservative a robust method is), that balances the decision robustness against the total system cost. The larger the $\mathit{\varGamma}$ or $\mathit{\alpha}$ the more lavish budget is needed to prepare for the risks. 

\begin{figure}
	\begin{center}
			\includegraphics[scale=0.5]{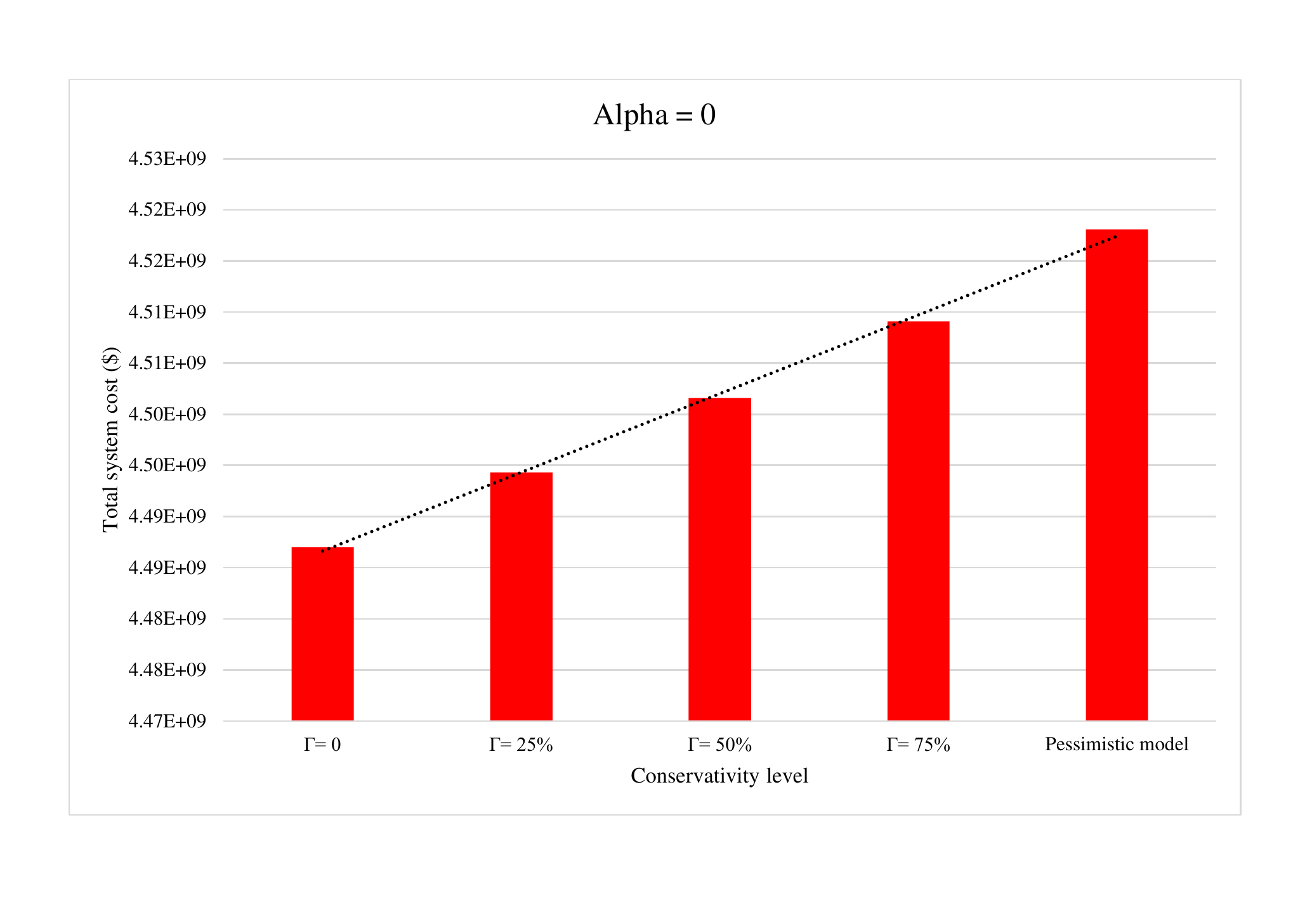} 
			\caption{Multi-stage adjustable Fuzzy robust approach to find the expected total system cost for \textit{$\mathit{\alpha}={0}$}}
		\label{figgg2}
	\end{center}
\end{figure}

\begin{figure}
	\begin{center}
			\includegraphics[scale=0.5]{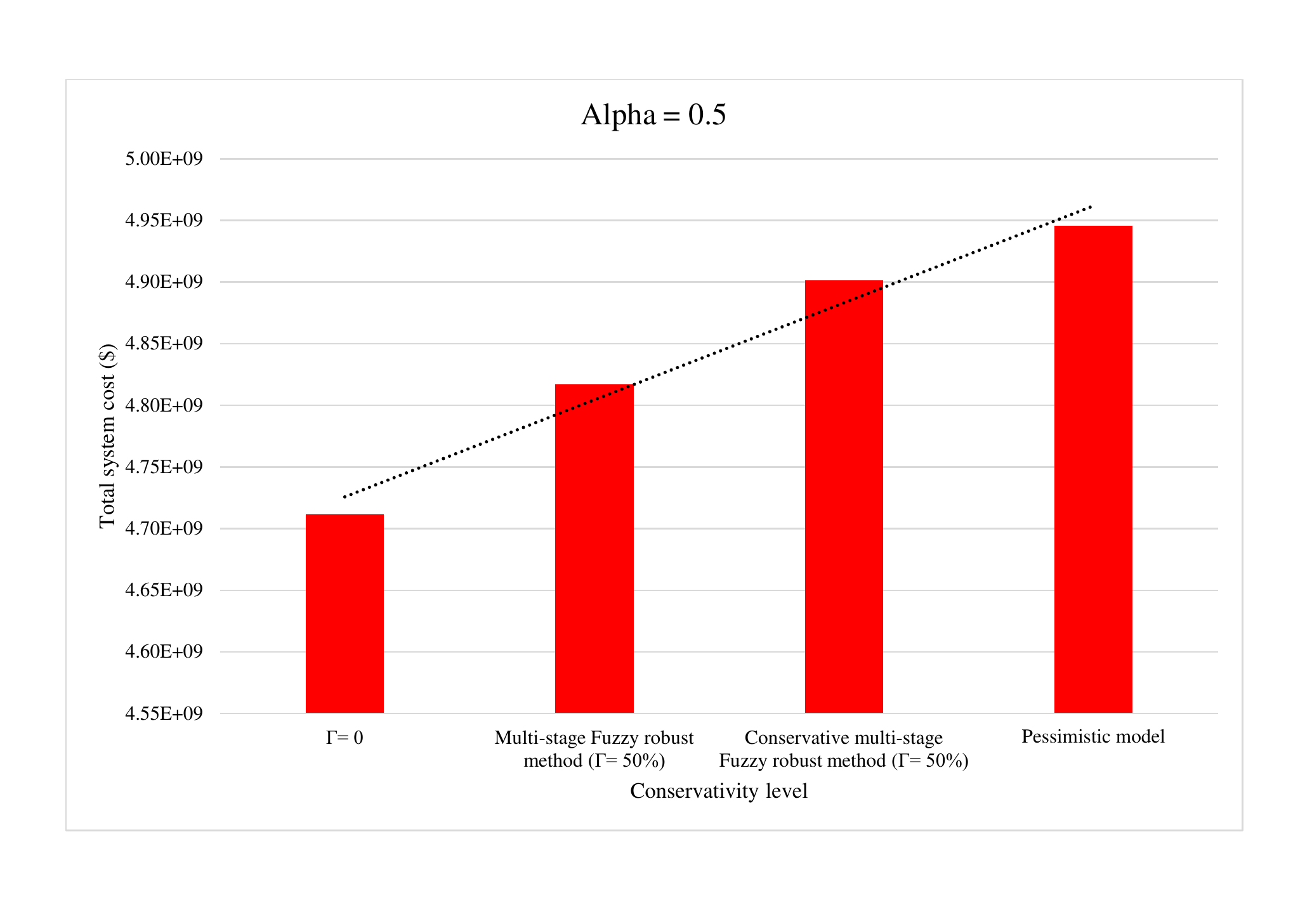}
		\caption{Multi-stage adjustable Fuzzy robust approach to find the expected total system cost for \textit{$\mathit{\alpha}={0.5}$}}
		\label{figgg3}
	\end{center}
\end{figure}

In the first numerical example (see Figure~\ref{figgg2}), different levels of $\mathit{\Gamma}$ (where the \textit{$\mathit{\alpha}={0}$}) can be picked based on the budget restrictions. In the second experiment (see Figure~\ref{figgg3}) the \textit{$\mathit{\alpha}={0.5}$} and the parameter\textit{$\mathit{\Gamma}$} is used to control the consevativity levels.

The multi-stage Fuzzy robust method incurs a lower total cost than all the other methods besides the deterministic model, which has inadequate dependability for real-world situations and can show infeasible decisions and water or power outages. If all the uncertain data had been added simultaneously to the model, the total system cost would have increased unreasonably, and the outcomes would have become impractical.

\section{Discussion} \label{section4}
One of the main challenges of energy-water nexus planning is to find an efficient way to deal with the uncertainties of the parameters. There are different types of randomness in energy-water nexus models that each requires a specific method to address. Fuzzy logic and robust optimizations are two popular methods to solve this issue. In this work, a new multi-stage adjustable Fuzzy robust approach is proposed to generate flexible, affordable, and efficient, reliable decisions without unreasonable cost-load on the total system cost.

The authors' proposed a multi-stage adjustable Fuzzy robust decision approach built on the authors' previous work that has lots of benefits; (1) yields flexible decisions that can be adjusted based on the budget constraints, (2) operates great with ranged uncertainty using robust optimization, (3) considers policy-makers and managers views to apply mathematically to the deterministic model using the Fuzzy logic, and (4) brings more cost-effective decisions to the system, in comparison with traditional conservative models. For future studies, the goal is to integrate the concept of Z-number \cite{ZADEH20112923} with the currently Fuzzy robust developed model.

%
\bibliographystyle{splncs03_unsrt}
\bibliography{author}

\end{document}